\documentclass[twocolumn,showpacs,preprintnumbers,amsmath,amssymb]{revtex4}

\usepackage{graphicx}
\usepackage{dcolumn}
\usepackage{bm}

\begin{document}

\title{Exchange-correlation energy of a multicomponent two-dimensional electron gas}

\author{K. K\"arkk\"ainen, M. Koskinen, S.M. Reimann$^*$, and M. Manninen}
\affiliation{Department of Physics, University of Jyv\"askyl\"a,
FIN-40351 Jyv\"askyl\"a, Finland}
\affiliation{$^*$Mathematical Physics, Lund Institute of Technology, Lund, Sweden}

\date{\today}

\begin{abstract}
We discuss the exchange-correlation energy of a multicomponent (multi-valley) 
two-dimensional electron gas and show that an extension of the recent
parametrisation\cite{attaccalite2002} of the exchange-correlation 
energy by Attacalite {\it et al.} describes well also the
multicomponent system. We suggest a simple mass dependence of the 
correlation energy and apply it to study the phase diagram of the 
multicomponent 2D electron (or hole) gas.
The results show that even a small mass difference of the components 
(e.g. heavy and light holes) decreases the concentration of the 
lighter components already at relatively high densities.
\end{abstract}
\pacs{-}

\maketitle

\section{Introduction}

One of the most interesting discoveries in semiconductor heterostructures
was the formation of a two-dimensional (2D) electron
gas\cite{ando1982}. For example, this has set the starting point 
for building nanostructures such as quantum dots or wires 
(for reviews see \cite{chakraborty1999,reimann2002,wiel2003}). 
Many of the simplified models describing semiconductor nanostructures are
based on the assumption of a homogeneous electron gas.
The effects of the detailed band structure are included only by using 
an effective mass for the electrons and by screening the
electron-electron interaction with a static dielectric constant. 
For the conduction electrons of a direct semiconductor, like GaAs for example, 
this approach describes fairly well the underlying electronic
structure\cite{chakraborty1999,reimann2002}. 

In the following, we call such a normal 2D electron gas as
a {\it two-component gas}, the components being the spin-up and 
spin-down electrons. 
Similarly, a polarised electron gas is called a one-component Fermi gas.
(In the literature, the latter is sometimes referred to as 
``spinless fermions'').

Layered semiconductors, however, provide several structures where the simple
picture of an ideal (one- or two-component) two-dimensional electron
gas is likely to fail. 
In elemental semiconductors, silicon and germanium, the conduction
band minimum consists of four equivalent minima at nonzero values of 
the $k$-vector and, moreover, these minima have non-isotropic 
effective masses. The resulting conduction electron gas has an
``internal'' degeneracy of eight (two from spin and four from the four 
minima), and is a possible example of an eight-component Fermi gas. 
The correlations in such an electron gas are much more complicated and 
have been studied in connection with the 3D electron-hole plasma 
in semiconductors\cite{vashista1974}. 

The valence band maximum in most semiconductors is degenerate,
consisting of two bands corresponding to heavy and light holes. 
In 3D structures the hole gas will then consist of two kinds of
holes. In 2D structures there will be an energy separation 
between the band minima due to the fact that heavy holes can be more easily 
localised than the light hole. Nevertheless, if the mass difference is
not too large, both kinds of holes can coexist and the correlations
between them have to be considered.
The hole gas of most semiconductors is one example of a 
four-component Fermi gas.
Electron layers in semiconductor heterostructures provide yet another system
of multicomponent electron (or hole) gas, intensively studied 
theoretically\cite{neilson1993,sarma1994,alatalo1994,conti1996}.
Recently, electron addition spectra of vertical quantum dots based on
layered structures have also been measured\cite{austing1998}.

Density functional theory (DFT) in the local (spin) density approximation
(LSDA) has shown to be a useful starting point for studies of the ground
state properties of electrons in quantum dots\cite{reimann2002}. 
The 2D conduction electron gas of the semiconductor is then treated as 
an interacting electron gas. In the LSDA, the 
exchange-correlation energy is approximated by that of the ideal 2D electron
gas, the effective mass and the static dielectric constant appearing
only as scaling factors (for energy and distance). 
Applying the density functional method to the above mentioned 
multicomponent structures calls for a parametrisation of the  
exchange-correlation energy ($\epsilon_{xc}$) for the multicomponent 
system in question. 
Tanatar and Ceperley\cite{tanatar1989}
computed the one and two-component (polarised and non-polarised 2D 
electron gas) using the quantum Monte Carlo technique. 
More recently Attaccalite {\it et al.}\cite{attaccalite2002}
presented new Monte Carlo results for these systems.
They also derived an interpolated expression for the
exchange-correlation energy functional to be used in 
DFT calculations.
The most important difference to previous parametrisations like for
example the one by Tanatar and Ceperly\cite{tanatar1989} is that 
here the Monte Carlo-calculations are performed also at intermediate 
polarisations. The usual approximation to treat the polarisation
dependence of the correlation energy on the same footing as the
(exactly known) exchange part\cite{reimann2002} can thus be abandoned. 

For a multicomponent system the situation becomes more complicated and 
the transferability of $\epsilon_{xc}$ from one system to another is
not so obvious. The purpose of this paper is to estimate the 
exchange-correlation energy of a general multicomponent 2D fermi gas.
We will first discuss the general high and low density limits and
the effect of mass difference between the components.
We will then show that an analytic continuation of the parametrisation by
Attaccalite {\it et al.}\cite{attaccalite2002} 
fits well the existing many-body calculations for multicomponent systems.
As reference data we use the calculations of Conti and Senatore\cite{conti1996}
for a four-component 2D electron gas and the results of 
Apaja {\it et al.}\cite{apaja1997} for a 2D charged Bose gas which can be
viewed as an upper bound for the energy of a Fermi gas with an infinite
number of internal degrees of freedom.

Finally, we will study the phase diagram suggested for this energy
functional and show that already a small mass difference will 
decrease the concentration of the lighter-mass components.
When the density gets low enough, eventually all systems become
one-component, as predicted by the 
Attaccalite {\it et al.}\cite{attaccalite2002} 
parametrisation for the normal 2D electron gas.
 



\section{Exchange-Correlation energy: General discussion}

We consider a homogenous gas consisting of $\Lambda$ different kinds of fermions,
or components, as we call them.
For instance, the different spin states of the homogenous gas are treated as different
components.
For a normal electron gas $\Lambda=2$ and for a fully polarised gas $\Lambda=1$. 
We assume that all the components have the same
charge, but they may have different masses. 
The total density $n$ of the gas is
\begin{equation}
n=\sum_{i=1}^{\Lambda}n_i=n\sum_{i=1}^{\Lambda}\nu_i,
\end{equation}
where $n_i$ is the number density and $\nu_i=n_i/n$ is a dimensionless 
concentration of the component $i$.
Note that the concentrations are normalised as $\sum_{i}^{\Lambda}\nu_i=1$.
The density parameter $r_s=1/\sqrt{\pi n}$ is always related to the total 
density $n$ of the gas.
The kinetic energy per particle of the noninteracting gas becomes 
\begin{equation}
\epsilon_k(r_s,\{\nu_i\},\{m_i\})=\frac{\hbar^2}{r_s^2}\sum_{i=1}^{\Lambda}\frac{\nu_i^2}{m_i},
\label{ek1}
\end{equation}
where $m_i$ is the mass of the component $i$. 
The total exchange energy is the sum 
of exchange energies of the different components. The total exchange energy per particle is then
\begin{equation}
\epsilon_x(r_s,\{\nu_i\})=-\frac{e^2}{4\pi\varepsilon_0\epsilon} \frac{8}{3\pi r_s}\sum_{i=1}^{\Lambda}\nu_i^{3/2}.
\label{ex1}
\end{equation} 
Note that while the kinetic energy depends on the masses of the components, the
exchange energy is independent of the masses.
The correlation energy per particle is defined as the difference between the exact energy 
and the Hartree-Fock energy, $\epsilon_c=\epsilon_{tot}-(\epsilon_k+\epsilon_x)$.
In true semiconductor systems the different components can represent, for example,
different band minima. The exchange energy has then also components coming
from the exchange between electrons belonging to these different minima.
Generally such exchange interaction is found to be small\cite{beliaev1998,hada2003}.
In the spirit of the effective mass approximation 
we will completely neglect such effects here.


In order to study the multicomponent correlation energy we will first
consider the case where all the components have the same mass.
In this case the exchange correlation energy can be written
as a function of the density parameter $r_s$ and the concentrations
$\nu_i$. The symmetry requires that the energy functional is symmetric with
respect of the interchange of the different concentrations $\nu_i$,
i.e., it does not depend on the indexing.
We define numbers 
\begin{equation}
Z_\gamma=\sum_{i=1}^\Lambda \nu_i^\gamma,
\label{znumber}
\end{equation}
which by construction have the desired symmetry.
We notice that $Z_1=1$, $\epsilon_x\propto Z_{3/2}$ and
the kinetic energy $\epsilon_k\propto Z_2$ (for equal masses).
In general any function $g(\{ \nu_i\})$ with the desired symmetry property can be written 
as a function of the $Z_\gamma$-numbers, for example, as a series expansion
\begin{equation}
g(\{ \nu_i\})= f(\{ Z_{\gamma_i} \})= 
a_1 Z_{\gamma_1}+a_2 Z_{\gamma_2}+ a_{12}Z_{\gamma_1}Z_{\gamma_2} + \cdots,
\label{serie}
\end{equation}
where $a_i$ are constants and $\gamma_i$ can be noninteger.
One way to approach the correlation energy is to use the suggestion
of Wigner\cite{wigner1934} who interpolated the exchange-correlation energy of an 
electron gas between the two limiting cases, the pure exchange in the high
density limit and the energy of the Wigner crystal in the low density limit:
\begin{eqnarray}
\begin{array}{l}
\epsilon_{xc}(r_s,\{ \nu_i \},\{m_i\})=\epsilon_{WC}(r_s)\\
\\
+g(r_s,\{ \nu_i \},\{m_i\})[\epsilon_x(r_s,\{ \nu_i \})-\epsilon_{WC}(r_s)]
\end{array}
\label{wigner}
\end{eqnarray}
where $\epsilon_{WC}$ is the classical Madelung energy of the Wigner crystal, which
does not depend on the masses. The advantage
of this approach is that the mass dependence appears only in the interpolation
function $g$. Furthermore, if the masses are equal, the interpolation function $g$
can immediately be written as in Eq. (\ref{serie}).

In order to estimate the interpolating function $g$ one would need results of 
accurate many-body calculations for systems with different numbers of components,
and for several values of $r_s$ and the concentrations $\nu_i$. 
We are not in the position to perform such computations at this point but
use existing data for special cases. For the two-component system
Attaccalite {\it et al.}\cite{attaccalite2002} have performed Monte
Carlo calculations. They fitted an accurate interpolation formula to
the results which is valid for any $r_s$ and polarisation.

For the four-component system, Conti and Senatore\cite{conti1996} have presented Monte Carlo
results for several values of $r_s$ in the case where all the concentrations $\nu_i$
are equal. For systems with more components there exist no accurate data.
However, the limit of infinite components can be approximated by the results 
obtained for charged bosons\cite{apaja1997}. 
If all the concentrations are the same ($\nu_i=1/\Lambda$)
in the limit of infinitely 
many components, all the particles are at different internal state and the Pauli
exclusion principle can not prevent putting all the particles in the same 
orbital state. The symmetric boson state is then a legitime state for such a
fermion system and provides an upper bound for the energy.
However, the existing results beyond the one and two-component systems are still
quite limited for an accurate interpolation of the function $g$.

\subsection{Mass dependence}

As discussed above, the exchange energy is mass-independent and even
in the multi-component electron gas only the kinetic and the 
correlation energies will depend on the masses. 
If all the components have the same mass $M$, the kinetic energy of Eq. (\ref{ek1})
can be written as $\epsilon_k=\hbar^2 Z_2/(r_s^2 M)$. 
In the general case, the kinetic
energy can still be written in the same form by defining an
{\it average mass} $M$ as
\begin{equation}
\frac{1}{M}=\frac{1}{Z_2}\sum_{i}^{\Lambda}\frac{\nu_i^2}{m_i}.
\end{equation}
We notice a scaling: By deviding the the kinetic and exchange energies by $M$, the resulting ratios
will only depend on the product $Mr_s$ 
($\epsilon_k/M=\hbar^2Z_2/(Mr_s)^2$ and $\epsilon_x/M=-8e^2Z_{3/2}/(4\pi\varepsilon_0\epsilon 3\pi Mr_s)$).
Moreover, considering the atomic units in the case that all masses are the same, 
we notice that (in this case) also the {\it total} energy, and thus
the correlation energy has the same scaling. It is then a reasonable first 
approximation to take the mass as an average mass and use it as a scaling factor
\begin{equation}
\epsilon_{xc}(r_s,\{ \nu_i \},\{ m_i \})=\frac{M}{m_e}\epsilon_{xc}(Mr_s,\{ \nu_i\},\{ m_i=m_e\})~,
\label{scale}
\end{equation}
where $m_e$ is the bare mass of the electron (or any suitably chosen effective mass).
(Note that within this approximation, the interpolation function $g$ in Eq. (\ref{wigner})
would be written as $g(Mr_s,\{ Z_\gamma \})$). 

\section{Extension of the two-component function}

For estimating an interpolation formula for the multicomponent 
correlation energy, we start by considering the 
parametrised form of the two-component 2D electron gas by 
Attaccalite \emph{et al.} \cite{attaccalite2002}, in order to see if their function
can be written in terms of the $r_s$ and the $Z_\gamma$-numbers, as discussed above. 
Attaccalite \emph{et al.} use the common description where the energy is written as
a function of $r_s$ and the polarisation $\zeta=(n_1-n_2)/n=\nu_1-\nu_2$
of the electron gas. They chose a correlation energy function 
$e_c(r_s,\zeta)=\epsilon_{xc}(r_s,\zeta)-\epsilon_x(r_s,\zeta)$
which diminishes the contributions of $\epsilon_x$ beyond 
fourth order in $\zeta$ as $r_s$ increases.
The parametrisation satisfies
known high- and low-density limits: For high densities i.e. small $r_s$ the exchange-correlation energy
obeys the perturbation theory result
$\epsilon_{xc}=\epsilon_x+a_0(\zeta)+b_0(\zeta)r_s\ln r_s$. For low densities 
the Wigner crystal limit is recovered, $\epsilon_{xc} \sim -c_1/r_s+c_2/r_s^{3/2}$, 
where the constants $c_1$ and $c_2$ are independent of the number of components.

Using the concentrations $\nu_i$, the square of the 
polarisation $\zeta$ of the two-component gas
becomes $\zeta^2=2(\nu_1^2+\nu_2^2)-1=2Z_2-1$. 
Now the exchange-correlation energy introduced
by Attacalite {\it et al.}\cite{attaccalite2002} can be 
written as (using their notations except for the above replacement for $\zeta$)
\begin{eqnarray} 
\label{eq:exc}
\begin{array}{l}
\epsilon_{xc}(r_s,\{\nu_i\})=e^{-\beta r_s}[\epsilon_x-\epsilon_x^{(6)}]\\
\\
+\epsilon_x^{(6)}+\alpha_0(r_s)
	+\alpha_1(r_s)(2Z_2-1)+\alpha_2(r_s)(2Z_2-1)^2~,
\end{array}
\end{eqnarray}
where $\epsilon_x^{(6)}=(1+\frac{3}{8}(2Z_2-1)+\frac{3}{128}(2Z_2-1)^2)\epsilon_x(r_s,\zeta=0)$.
The functions $\alpha_i(r_s)$ are parametrised by Attacalite {\it et
  al.}\cite{attaccalite2002} and are independent of the concentrations.
This two-component exchange-correlation function of Attaccalite {\it et al} is thus already of the 
desired form: It depends only on $r_s$, $Z_2$ and $Z_{3/2}$ (through
$\epsilon_x$) and can be directly extended to any number of components.

For a given value of $r_s$  Attaccalite {\it et al}\cite{attaccalite2002} 
fitted their function to several values
of polarisations, $\zeta \in [0,1]$, which correspond to $Z_2\in [0.5,1]$ and $Z_{3/2}\in [\sqrt{2},1]$.
An analytic continuation to larger number of components means an extrapolation of the $\epsilon_{xc}$
to the region $Z_2\in [0,1]$ and $Z_{3/2}\in [0,1]$ for a system with 
infinitely many components,
but only to  $Z_2\in [0.25,1]$ and $Z_{3/2}\in [0.5,1]$ for a four-component system.
Moreover, since $Z_{3/2}$ only appears in the exchange energy, Eq. (\ref{ex1}), which is exact for any
number of components,
errors in the extrapolation only result from the extension of the $Z_2$ space.

The multicomponent gas must obey the 
low density Wigner crystal limit and the high density exchange limit; both of these
requirements are satisfied by Eq. (\ref{eq:exc}). 
By taking a ``paramagnetic'' gas, 
i.e. by setting $\nu_i=\Lambda^{-1}$ for all the components and fixing $r_s$,
it can be seen that the exchange-correlation energy above is also a monotonous function
of the number of components. 

\begin{figure}
\includegraphics[angle=-90]{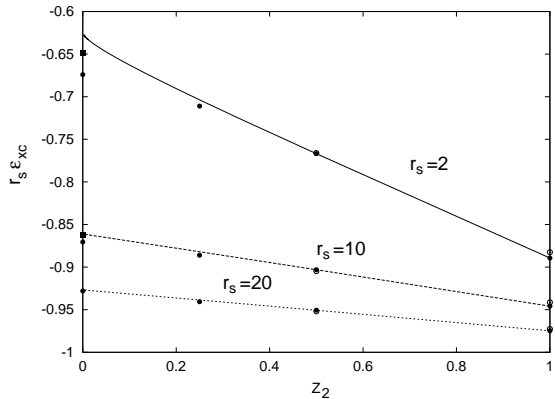}
\caption{Dependence of the exchange-correlation energy of
the multicomponent electron gas as a function of parameter $Z_2$,
defined by Eq. (\ref{znumber}).
The lines show the result of the interpolation formula of
Attaccalite {\it at al.}\cite{attaccalite2002}, Eq. (\ref{eq:exc}), 
extended to multicomponent systems.
The points correspond to numerical results of many-body calculations as follows:
Black dots for $Z_2=1$ and $Z_2=0.5$ are from Ref.\cite{attaccalite2002};
black dots for $Z_2=0.25$ from Ref.\cite{conti1996}; black dots for $Z_2=0$ from
Ref.\cite{apaja1997}; open circles from Ref.\cite{tanatar1989}; black squares from
Ref.\cite{gold1992}. Results for $r_s=2$, 10, and 20 are shown.}
\label{fig1}
\end{figure}

Figure 1 shows the exchange-correlation energy derived from Eq. (\ref{eq:exc}) for selected 
values of $r_s$ as a function of $Z_2$. For the interval $Z_2\in [.5,1]$ the lines correspond to 
the normal two-component electron gas with partial polarisation. For
the 
interval $Z_2\in [0,.5]$
the lines are derived for the case where all the concentrations are equal, $\nu_i=1/\Lambda$
and consequently $Z_2=1/\Lambda$. The points are results of different many-body calculations.
Note that the lines from $Z_2=0.5$ to $Z_2=1$ agree exactly to the results of 
Attaccalite {\it et al}\cite{attaccalite2002}. We can see that the extension to a four component
system ($Z_2=.25$) is fairly accurate and the extension even to the boson case ($Z_2=0$) is still 
reasonable. In fact,
the agreement of the Attaccalite {\it et al.}-function with the existing
many-body results for four- and infinite-component systems is so good that there is no need to
make a better interpolation formula for the multicomponent system, before more total energy
calculations are available in the region $Z_2<.5$. It would be 
important to have results also 
for cases where the concentrations of all the components are not equal.

\section{Phase diagram of the multicomponent 2D electron gas}

It was recently suggested by Attacalite {\it et al.}\cite{attaccalite2002}
that the 2D electron gas shows a transition to a 
polarised gas at $r_s\approx 25$\cite{attaccalite2002}.
We will first study the stability of the multicomponent gas assuming all the 
masses to be equal. The extended exchange-correlation energy functional 
suggests that all multicomponent cases have a transition to a
one-component phase when $r_s$ increases. 
Table I gives the estimated transition points.
For any number of components, the transition happens nearly at the same point 
and directly from the multicomponent to a one-component phase. 
It is caused by the slightly
different $r_s$-dependence of the one-component total energy as compared to the others.
For $\Lambda \geq 2$ the total energy is a monotonously decreasing function of $\Lambda$
(for any $r_s$).

\begin{table}
\caption{Estimated values of $r_s$ where a $\Lambda$-component system 
spontaneously transforms to a one-component system, assuming that all the 
components have the same mass.}

\medskip

\begin{tabular}{llllll}\hline\hline
$ \Lambda$  \qquad\qquad\qquad&  2 \qquad \qquad   &  4 \qquad  \qquad  &
  6  \qquad  \qquad & 8 \qquad \qquad & $\infty $  \qquad \qquad \\ \hline
$r_s$      & 25.6 & 26.2 & 26.4 & 26.6 & 27.0    \\
\hline\hline
\end{tabular}
\end{table}

\medskip


Next we will study the effect of the mass difference, using the approximation of 
Eq. (\ref{scale}). We assume a four-component system and
fix the masses pairwise equal so that two components are heavier than the two others,
say  $m_1=m_2=m_l$ and  $m_3=m_4=m_h$.
Note that due to the spin degeneracy of electrons (or holes) there will always
be two components with the same mass, and the number of components will be even, except
of the special case one. 
The phase diagram of the four-component electron gas obtained using Eq. (\ref{eq:exc}) is shown in 
figure 2. We show the maximum of the concentrations ($\nu_4$) as
a function of $r_s$ and the mass ratio $m=m_h/m_l$. 
When the masses are equal, also the concentrations remain equal ($\nu_i=0.25$) until
$r_s\approx 26.2$ at which point the system becomes a one-component gas.
When the mass ratio increases, the concentrations $\nu_3$ and $\nu_4$ increase due to the 
reduction of the kinetic energy, and eventually $\nu_3=\nu_4=0.5$ and the system has become
two-component. A further increase of the mass ratio eventually changes the 
system to a one-component gas. We notice that
as the mass is increased the polarisation occurs at smaller $r_s$. 
The reason is the scaling of the total energy and the dimensions with the mass.

\begin{figure}
\includegraphics[angle=0]{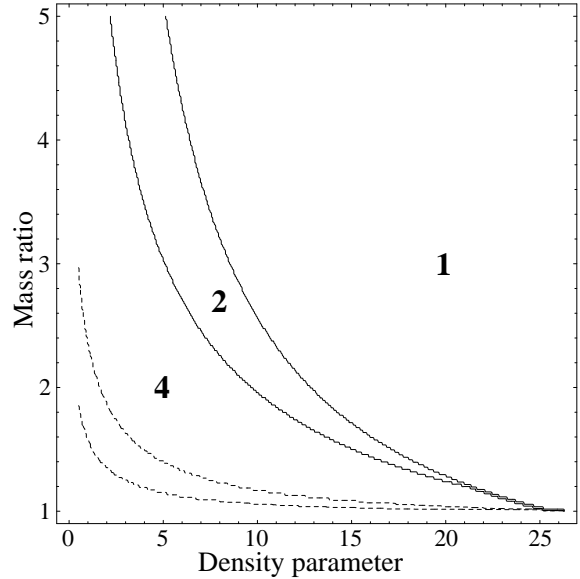}
\caption{Phase diagram for the four-component gas in the $r_s$-$m_h/m_l$ plane.
The solid lines separate the 1, 2 and 4-component phases (denoted bu numbers).
In the four-component phase the concentrations of the components 
with the heavy mass, say $\nu_3$ and $\nu_4$,
decrease continuosly from 0.5 to 0.25 in going from the lowest solid line to
the mass ratio 1 or to $r_s=0$. The dashed lines correspond to concentrations
$\nu_3=\nu_4=0.42$ and $\nu_3=\nu_4=0.35$.
}
\label{fig2}
\end{figure}

In real semiconductor heterostructures the mass difference can arise from different
constituents of the two layers in a double layer system or, in the case of holes, simply
from the mass difference of the heavy and light hole. In both cases, however,
there can be a constant potential energy difference from one component to another.
This energy difference arises from the localisation of the particle in the direction
perpendicular to the 2D-layer. For example, if we consider holes in a 1D harmonic potential,
the ground state energy (the lowest perpendicular mode) has the energy
$\epsilon_0(m)=\hbar\omega_0/(2\sqrt{m/m_e})$, where $\omega_0$ is the confinement strength 
for $m_e$.
This energy is smaller for
the heavy hole and, consequently, further favours the transition to a two-component system
where only heavy holes exist. In order to have a 2D electron gas the Fermi energy has to 
be clearly below the first excited perpendicular state, 
$\epsilon_1(m)=3\hbar\omega_0/(2\sqrt{m/m_e})$ in a harmonic well.
In order for the heavy (mass $m_h$) and light holes (mass $m_l$) to coexist, the ratio
$(\epsilon_0(m_l)-\epsilon_0(m_h))/(\epsilon_1(m_h)-\epsilon_0(m_h))$ has to be clearly smaller than
one. For example in the cases of Si, Ge and GaAs ($m_h/m_l>5$) this
ratio is larger than 0.6. 
Consequently, a 2D layer with both heavy and light holes is not possible.
However, there are semiconductors, like AlAs, where $m_h/m_l\approx 2$ and the above energy 
ratio is only about 0.2, making the four-component gas possible.

\section{Conclusions}

We have studied the possibility to formulate a general
multicomponent 2D exchange-correlation energy to be used in
density functional calculations.
After general considerations of the concentration and mass dependence 
of the correlation energy, 
we have demonstrated that the recent 
parametrisation by Attacalite {\it et al.}\cite{attaccalite2002} 
of the two-component exchange-correlation energy 
has functional properties which allow a direct extension to multicomponent
systems. Furthermore, although the function is fitted only to the 
two-component electron gas,
it gives fairly accurate results for four-component and infinite-component
systems.

We suggest that the mass dependence of the correlation energy can be estimated with
a properly chosen average mass. The application for multicomponent 2D gas shows 
an interesting phase diagram. For equal masses all systems transform from a
multicomponent directly to a one-component gas when $r_s$ increases to about 
$25\cdots 27$ depending on the number of components.
Even a small mass difference decreases the concentration of the lighter component and it
transforms first to a purely two-component and then to a one-component system when $r_s$ increases.

A more accurate interpolation formula requires extensive many-body calculations for multi-component
gases. We hope that these considerations encourage such work.


\begin{thebibliography}{23}

\bibitem{attaccalite2002} C. Attaccalite, S. Moroni, P. Gori-Giorgi and G.B. Bachelet, Phys. Rev. Lett., 
	{\bf 88}, 256601 (2002).

\bibitem{ando1982} T. Ando, A.B. Fowler, and F. Stern, Rev. Mod. Phys. {\bf 54}, 437 (1982).

\bibitem{chakraborty1999} T. Chakraborty,
{\it Quantum Dots: A survey of the properties of artificial
atoms} (North-Holland, Amsterdam 1999).

\bibitem{reimann2002} S.M. Reimann and M. Manninen, Rev. Mod. Phys. {\bf 74}, 1283 (2002).

\bibitem{wiel2003}  W. G. van der Wiel, S. De Franceschi, J. M. Elzerman, T. Fujisawa, 
S. Tarucha, and L. P. Kouwenhoven, Rev. Mod. Phys. {\bf 75}, 1 (2003).

\bibitem{vashista1974} P. Vashista, P. Bhattaracharyya and K. S. Singwi, Phys. Rev. B 10, 5108 (1974). 

\bibitem{neilson1993} D. Neilson, L. \'Swierkovski, J. Szyma\'nski, and L. Liu,
Phys. Rev. Lett. {\bf 71}, 4035 (1993).

\bibitem{sarma1994} S. Das Sarma and P.I. Tamborenea, Phys. Rev. Lett. {\bf 73}, 1971 (1994).

\bibitem{alatalo1994} M. Alatalo, P. Pietil\"ainen, T. Chakraborty, and M. Salmi,
Phys. Rev. B {\bf 49}, 8277 (1994).

\bibitem{conti1996} S. Conti and G. Senatore, Europhys. Lett., {\bf 36}, 695 (1996).

\bibitem{austing1998} D.G. Austing, T. Honda, K. Muraki, Y. Tokura, and
S. Tarucha, Physica B {\bf 249-251}, 206 (1998).

\bibitem{tanatar1989} B. Tanatar and D.M. Ceperley, Phys. Rev. B, {\bf 39}, 5005 (1989).

\bibitem{apaja1997} V. Apaja, J. Halinen, V. Halonen, E. Krotscheck, and M. Saarela,
Phys. Rev. B, {\bf 55}, 12925 (1997).

\bibitem{wigner1934} E. Wigner, Phys. Rev. {\bf 46}, 1002 (1934).

\bibitem{beliaev1998} D. Beliaev, L.M. Scolfaro, J. R. Leite, and G.M. Sipahi,
Phys. Stat. Sol. (b) {\bf 210}, 777 (1998).

\bibitem{hada2003} Y. Hada and M. Eto, cond-mat/0304228 (2003).

\bibitem{gold1992} A. Gold, Z. Phys. B {\bf 89}, 1 (1992)








\end{thebibliography}
\end{document}